\documentclass[a4paper,fleqn,usenatbib]{mnras}

\usepackage{newtxtext,newtxmath}
\usepackage[T1]{fontenc}
\usepackage{ae,aecompl}
\usepackage{hyperref}
\usepackage[dvipdfmx]{graphicx}	
\usepackage{amsmath}	
\usepackage{amssymb}	
\usepackage{bm}

\title[Radio Detectability of Isolated Black Holes]{Radio Emission from Accreting Isolated Black Holes in Our Galaxy}
\author[D.Tsuna et al.]{
Daichi Tsuna$^{1,2}$\thanks{tsuna@resceu.s.u-tokyo.ac.jp}
and Norita Kawanaka$^{3,4}$\\
$^{1}$Research Center for the Early Universe (RESCEU), the University of Tokyo, Hongo, Tokyo 113-0033, Japan\\
$^{2}$Department of Physics, School of Science, the University of Tokyo, Hongo, Tokyo 113-0033, Japan\\
$^{3}$Department of Astronomy, Graduate Sool of Science, Kyoto University, Kitashirakawa Oiwake-cho, Sakyo-ku Kyoto, 606-8502, Japan\\
$^{4}$Hakubi Center, Yoshida Honmachi, Sakyo-ku, Kyoto 606-8501, Japan}
\date{Accepted XXX. Received YYY; in original form ZZZ}
\pubyear{2019}

\begin{document}
\label{firstpage}
\pagerange{\pageref{firstpage}--\pageref{lastpage}}
\maketitle

\begin{abstract}
Apart from the few tens of stellar-mass black holes discovered in binary systems, an order of $10^8$ isolated black holes (IBHs) are believed to be lurking in our Galaxy. Although some IBHs are able to accrete matter from the interstellar medium, the accretion flow is usually weak and thus radiatively inefficient, which results in significant material outflow. We study electron acceleration generated by the shock formed between this outflow and the surrounding material, and the subsequent radio synchrotron emission from accelerated electrons.  By numerically calculating orbits of IBHs to obtain their spatial and velocity distributions, we estimate the number of IBHs detectable by surveys using SKA1-mid (SKA2) as $\sim 30$ ($\sim 700$) for the most optimistic case. The SKA's parallax measurements may accurately give their distances, possibly shedding light on the properties of the black holes in our Galaxy.
\end{abstract}


\section{Introduction}

Since the first observations of black holes in the previous century, a few tens of stellar-mass black holes have been detected via X-rays (for a review see \citealt{RM06}) and gravitational waves \citep{GW150914,GW151226,GW170104,GW170814,GW170608,GWTC-1}. These observations have given us important clues on their properties (e.g. mass, spin) and have helped push the frontiers of stellar evolution and accretion disc theory. However, these stellar-mass black holes are only a very tiny fraction of the black holes that exist in the universe, or even in our Galaxy. In fact the known stellar-mass black holes in the Galaxy are all binaries, and it is believed that there exists a large population of isolated black holes (IBHs) without a companion star. Past studies based on models of e.g. stellar evolution or chemical evolution in the Galaxy predict that the number of black holes formed in the history of the Galaxy is of order $10^8$ \citep{Shapiro83,vandenHeuvel92,Samland98,Caputo17}. The fraction of IBHs among them should be non-negligible, and may even occupy a majority \citep{Fender13}.

A naive way to observe IBHs is through their X-ray emission via accretion of gas in the interstellar medium. Past theoretical works regarding the detectability of IBHs have thus mainly focused on accreting sources \citep{Shvartsman71,McDowell85,CP93,PP98,Fujita98,Armitage99,Agol02,Maccarone05,MT05,Barkov12,Fender13,Ioka17,Matsumoto17,Tsuna18}. Although past searches using X-ray data have found candidates of accreting IBHs \citep{Chisholm03,Muno06}, no strong candidates are known at present. This is probably not only because of the lower accretion rate compared to X-ray binaries, but also because of the lower radiative efficiency. The accretion flow for low radiative efficiency is known as radiatively inefficient accretion flow \citep[RIAF;][]{Ichimaru77,NY95}. At this state, it is predicted that much of the accreted material is swept away as outflows, and that only a small fraction can reach near the black hole to shine in X-rays \citep{BB99,Narayan00,Quataert00}.

These outflows can possibly make the IBHs detectable in other wavelengths. The outflows can interact with the surrounding matter and create strong collisionless shocks at the interface. These shocks can amplify magnetic fields and accelerate electrons, and these electrons emit synchrotron radiation in the radio wavelength.

Radio emission from IBHs and their detectability had been considered previously in \cite{Maccarone05} and \cite{Fender13}. These works obtained radio luminosities of the IBHs by utilizing the radio/X-ray luminosity correlation $L_R\propto L_X^{0.7}$ known for black hole binaries in the low/hard state \citep{Corbel03,Gallo03,Gallo06,Corbel13}. The correlation has been explained by kinetic energy dissipation of a jet being launched from the black hole \citep{Fender01,Koerding06}, but some radio-quiet population that deviate from this correlation is known \citep{Coriat11,Gallo12}. The radio emission mechanism proposed in this paper is more general and do not rely on the extrapolation of empirical radio/X-ray correlation down to the dimmest luminosities yet to be probed by observations.

Radio astronomy is expected to be revolutionized by the upcoming Square Kilometer Array (SKA), whose expected sensitivity surpasses existing radio surveys by orders of magnitude\footnote{\url{https://astronomers.skatelescope.org/ska/}}. SKA is planned to be constructed in two phases: phase 1 (SKA1), and phase 2 (SKA2). The covered frequency is from 50 MHz to 15 GHz, with its highest sensitivity in the GHz regime. We propose in this paper that IBHs can be one of the promising targets for SKA, and aim to give an estimate on the number of detectable IBHs in the whole Galaxy by the two SKA phases.

This paper is organized as follows. In Section \ref{sec:formulations}, our model of radio emission and the setup of our calculations are presented. In Section \ref{sec:results} we show our prospects for detecting these radio emission by SKA1 and SKA2. We discuss in Section \ref{sec:discussion} the possibility of obtaining the properties of the detectable IBHs from radio and multi-wavelength observations, as well as possible caveats in our calculations. We conclude in Section \ref{sec:conclusions}.


\section{Formulations}
\label{sec:formulations}

\subsection{RIAF Accretion and outflow}
Let us consider a black hole of mass $M_{\rm BH}$ and velocity $\upsilon_{\rm BH}$, plunging into a gas cloud of density $\rho$ and sound speed $c_{\rm ISM}$. The accretion rate in the outermost region can be approximated by Bondi's formula \citep{Bondi52}:
\begin{align}
\dot{M}_B &\approx4\pi \frac{(GM_{\rm BH})^2}{\upsilon^3} \rho\nonumber \\
		&\sim 3.7\times 10^{16}\ {\rm g\ s^{-1}} \left(\frac{\upsilon}{10\ {\rm km\ s^{-1}}}\right)^{-3} \left(\frac{M_{\rm BH}}{10\ M_\odot}\right)^2 \left(\frac{\rho}{10^3\ \mathrm{cm^{-3}}\ m_p}\right),
\end{align}
where we have defined $\upsilon=(\upsilon_{\rm BH}^2+c_{\rm ISM}^2)^{1/2}$, and $G$ and $m_p$ are the gravitational constant and proton mass respectively. This accretion rate is much smaller than the Eddington accretion rate
\begin{align}
\dot{M}_{\rm Edd} \approx 1.4\times 10^{19}\ {\rm g\ s^{-1}} \left(\frac{M_{\rm BH}}{10\ M_\odot}\right) \left(\frac{\eta_{\rm std}}{0.1}\right)^{-1},
\end{align}
where $\eta_{\rm std}$ is the radiation efficiency of the standard disc. 

When the mass accretion rate is much smaller than $\dot{M}_{\rm Edd}$, the flow will be represented by the so-called radiatively-inefficient accretion flow (RIAF), where radiative cooling will not be efficient and the heat dissipated by viscosity is transported inwards by advection \citep{Ichimaru77,NY95,KFM08}. \citet{NY95} have found that the transition from a standard disc \citep{Shakura73} to a RIAF depends on the strength of the viscosity in the accretion flow. For the viscosity parameter $\alpha = 0.1$--$0.3$, this transition is estimated to be around $10^{-2}$--$10^{-1}{M}_{\rm Edd}$ \citep{Narayan05,KFM08}. In this work we assume the threshold is at $10^{-1}{M}_{\rm Edd}$.

In the RIAF regime the internal energy advected inward in the accretion flow can launch an outflow from the inner disc region, which suppresses the accretion on to the central black hole from the Bondi accretion rate. This is considered to be the case for Sgr A*, whose bolometric luminosity is observed \citep{Baganoff03} to be much smaller than that inferred from the standard disc model assuming Bondi accretion rate. 

Outflows for RIAFs are speculated to occur due to the gas in the accretion flow being unbound \citep{NY95,BB99}, and/or by magnetic field amplification inside the disc that creates turbulence and high magnetic pressure. Hydrodynamical and MHD simulations support that either or both of these two processes make significant contributions in the production of outflows \citep[see][for a review]{Perna03,KFM08}.

One of the representations for this accretion flow is the self-similar solution obtained by \cite{BB99}, where they assumed that the accretion rate scales with radius as $\dot{M}(r) \propto r^{p}$. Some works indicate that RIAF can become convectively unstable, due to the increase in gas entropy as heat is advected to the centre \citep{Narayan00,Quataert00}. These works have found a solution known as a convection-dominated accretion flow (CDAF), where the power-law index becomes $p\approx 1$.

\begin{figure}
 \begin{center}
  \includegraphics[width=1.1\linewidth]{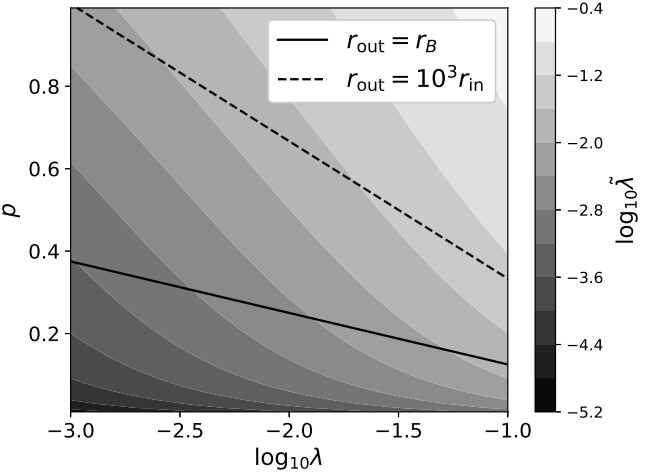} 
 \end{center}
\caption{Values of $\tilde{\lambda}$, defined in equation \eqref{eq:lambda_tilde}, as a function of $p$ and $\lambda$. The solid line shows the case where the accretion disc's outer edge radius $r_{\rm out}$ is equal to the Bondi radius $r_B$. The dashed line shows the case where the accretion disc's outer edge radius is $10^3$ times the inner edge radius $r_{\rm in}$.}
\label{fig:lambda_tilde_contour}
\end{figure}

The properties of the outflow can be estimated by assuming the radial profile of the outflow rate and the velocity of the outflow. We assume that the radial profile of the accretion rate is a power-law following previous works, and that $\dot{M}=\dot{M}_B$ at the outer edge of the RIAF $r=r_{\rm out}$. The matter inflow rate as a function of radius is then
\begin{align}
\dot{M} = \dot{M}_B \left(\frac{r}{r_{\rm out}}\right)^p.
\end{align}
We further assume that the velocity of the outflow is similar to the escape velocity at the radius where it is launched $\upsilon_{\rm esc} \sim \sqrt{2GM_{\rm BH}/r} \propto r^{-1/2}$. The outflow rate and power becomes
\begin{align}
\dot{M}_{\rm out} \sim \int_{r_{\rm in}}^{r_{\rm out}}\frac{\mathrm{d}\dot{M}}{\mathrm{d}r} \mathrm{d}r \sim \dot{M}_B \left[1-\left(\frac{r_{\rm in}}{r_{\rm out}}\right)^{p}\right] \equiv (1-\lambda) \dot{M}_B,
\end{align}
and
\begin{align}
L_{\rm out} &= \int_{r_{\rm in}}^{r_{\rm out}} \frac{1}{2} \frac{\mathrm{d}\dot{M}}{\mathrm{d}r} \upsilon_{\rm esc}^2 \mathrm{d}r =  \frac{p(\lambda-\lambda^{1/p})}{1-p}\dot{M}_B\frac{GM_{\rm BH}}{r_{\rm in}}.
\end{align}
Here $\lambda \equiv (r_{\rm in}/r_{\rm out})^p \ll 1$ corresponds to the fraction of accreted matter that actually reaches the BH. The values of $\lambda$ and $p$ are uncertain, both theoretically and observationally. Observations of nearby active galaxies give a typical value of around $\lambda\sim 0.01$ \citep{Pellegrini05}. The small number of X-ray detections of accreting neutron stars by {\it ROSAT} gives a limit of $\lambda\lesssim 10^{-3}$ \citep{Perna03}, but the situation can be different for black holes that do not have strong magnetic fields or rigid surfaces. For the power-law index $p$, hydrodynamical and MHD simulations agree with a value of $p=0.5$--$1$ \citep[see][]{Perna03}. However a fit to the observed spectral energy distribution of Sgr A* prefers a smaller a value of $p=0.27$ \citep{Yuan03}. In this work we assume $\lambda$ and $p$ in the range $10^{-3}<\lambda<0.1$ and $0<p<1$. The outflow power is calculated as a function of $\lambda$ and $p$ as
\begin{align}
L_{\rm out} &\approx  1.7\times 10^{35}\ {\rm erg\ s^{-1}} \nonumber \\
&  \left(\frac{\tilde{\lambda}}{0.03} \right) \left(\frac{\upsilon}{10\ {\rm km\ s^{-1}}}\right)^{-3} \left(\frac{M_{\rm BH}}{10\ M_\odot}\right)^2 \left(\frac{\rho}{10^3\ \mathrm{cm^{-3}}\ m_p}\right) \left(\frac{r_{\rm in}/r_{\rm S}}{3}\right)^{-1},
\label{eq:Ljet}
\end{align}
where
\begin{align}
\tilde{\lambda} \equiv \left[\frac{p(\lambda-\lambda^{1/p})}{1-p}\right],
\label{eq:lambda_tilde}
\end{align}
and $r_{\rm S}$ is the BH's Schwarzschild radius. A profile of the parameter $\tilde{\lambda}$ is plotted in Figure \ref{fig:lambda_tilde_contour} as a function of $\lambda$ and $p$. For our simple accretion disc model, we can consider the range of values that $\tilde{\lambda}$ can take. A natural condition that $r_{\rm out}<r_B$, where
\begin{align}
r_B\sim GM_{\rm BH}/\upsilon^2 \sim 1.3\times 10^{15}\ {\rm cm} \left(\frac{\upsilon}{10\ {\rm km\ s^{-1}}}\right)^{-2} \left(\frac{M_{\rm BH}}{10\ M_\odot}\right)
\end{align}
is the Bondi radius, imposes a constraint on $\tilde{\lambda}$ as being above the black solid line in Figure \ref{fig:lambda_tilde_contour}. This leads to $\tilde{\lambda} \gtrsim 2\times 10^{-3}$. Furthermore, the value of $\tilde{\lambda}$ for the case where $r_{\rm out}=10^3r_{\rm in}$ is shown in Figure \ref{fig:lambda_tilde_contour} as a dotted line. In this case, $\tilde{\lambda}$ takes values of $10^{-2} \lesssim \tilde{\lambda} \lesssim  5\times 10^{-2}$.

The typical outflow velocity is
\begin{eqnarray}
\upsilon_{\rm out} \sim \sqrt{\frac{2L_{\rm out}}{\dot{M}_{\rm out}}} \sim 0.1c  \left(\frac{\tilde{\lambda}/(1-\lambda)}{0.03} \right)^{1/2} \left(\frac{r_{\rm in}/r_{\rm S}}{3}\right)^{-1/2},
\end{eqnarray}
which is scale-free if $p$, $\lambda$ and $r_{\rm in}/r_{\rm S}$ do not depend on the size of the BH. For the rest of our work we approximate as $1-\lambda \approx 1$ to reduce the extra dependence on $\lambda$. This will lead to a slightly more conservative estimate on $\upsilon_{\rm out}$ and the radio emission.

\subsection{Shock Formation and Electron Acceleration}
The outflow collides with the interstellar medium and forms a shock at the interface, as $\upsilon_{\rm out}$ is much larger than the sound speed inside the ISM, $c_{\rm ISM} \lesssim 100~{\rm km}~{\rm s}^{-1}$. The particle number density in the ISM is low enough that the shock will be collisionless. In this case, magnetic fields are amplified, and electrons are shock-accelerated through the diffusive shock acceleration mechanism \citep[for a review see e.g.][]{BE87}.

The location of the termination shock corresponds to where the outflow's ram pressure balances the ISM inertia \citep{Barkov12}. Assuming the outflow is isotropic, its radius is 
\begin{align}
R_{\rm TS} & \sim \sqrt{\frac{2L_{\rm out}}{\rho \upsilon^2 4\pi \upsilon_{\rm out}}} \nonumber \\
  &\sim 7.1\times 10^{16}\ {\rm cm} \nonumber \\
  & \left(\frac{\tilde{\lambda}}{0.03} \right)^{1/4} \left(\frac{\upsilon}{10\ {\rm km\ s^{-1}}}\right)^{-5/2} \left(\frac{M_{\rm BH}}{10\ M_\odot}\right) \left(\frac{r_{\rm in}/r_{\rm S}}{3}\right)^{-1/4},
\end{align}
which is generally larger than the Bondi radius. 

Let us estimate the strength of the amplified magnetic field and the amount of non-thermal electrons accelerated at the terminal shock. We consider a stable one-zone system where a fraction $\epsilon_B$ of the internal energy density at the shocked region is transferred to the magnetic field, and a fraction $\epsilon_e$ is used to accelerate the electrons to relativistic energies. We can write
\begin{align}
V_{\rm sh} \frac{B^2}{8\pi} &\sim \epsilon_B L_{\rm out}  \frac{R_{\rm TS}}{\upsilon_{\rm out}} \label{eq:eps_B}\\
V_{\rm sh} \langle e_e \rangle &\sim \epsilon_e L_{\rm out}  \frac{R_{\rm TS}}{\upsilon_{\rm out}} \label{eq:eps_e},
\end{align}
where $B$ is the magnetic field strength, $V_{\rm sh}$ is the volume of the shocked region, and $\langle e_e \rangle$ is the average energy density of the accelerated electrons. We assume that the volume of the shocked region is comparable to the volume enclosed within radius $R_{\rm TS}$, i.e. $V_{\rm sh}\sim 4\pi R_{\rm TS}^3/3$. Then equation \eqref{eq:eps_B} gives
\begin{align}
B &\sim \ 79{\rm \mu G} \left(\frac{\epsilon_B}{0.1}\right)^{1/2} \left(\frac{\rho}{10^3\ \mathrm{cm^{-3}}\ m_p}\right)^{1/2} \left(\frac{\upsilon}{10\ {\rm km\ s^{-1}}}\right).
\label{eq:magnetic_field}
\end{align}
Assuming a power-law spectrum for the accelerated electrons as $n(\gamma_e)\mathrm{d} \gamma_e=n_0 \gamma_e^{-q} \mathrm{d} \gamma_e\ (\gamma_{\rm min} < \gamma < \gamma_{\rm max}, 2<q<3)$, we can estimate $n_0$ by setting $\gamma_{\rm min}$ and $\gamma_{\rm max}$ and solving equation \eqref{eq:eps_e}. Recent particle-in-cell simulations of diffusive shock acceleration give an estimate on the minimum Lorentz factor of the non-thermal electrons as 
\begin{align}
\gamma_{\rm min} -1 = \frac{\zeta_e}{2} \frac{m_p}{m_e}\left(\frac{\upsilon_{\rm out}}{c}\right)^2 \sim 3.7 \left(\frac{\zeta_e}{0.4}\right) \left(\frac{\upsilon_{\rm out}}{0.1c}\right)^2,
\end{align}
where $\zeta_e \approx 0.4$ \citep{Park15} is the efficiency of the kinetic energy of protons being transferred to electrons. Hereafter we use the value $\zeta_e = 0.4$, as was done in other works that considered radio transients by mildly relativistic outflows \citep{Murase16,Kimura17,Kashiyama18}. We note that this estimate on $\gamma_{\rm min}$ may be inaccurate for outflow velocities much lower than $\sim 0.1c$. The maximum Lorentz factor of the electrons is given by the condition that the acceleration timescale must be smaller than the dynamical timescale and the synchrotron timescale. Each timescale is estimated to be
\begin{align}
t_{\rm acc} &\sim  \frac{20\eta}{3} \frac{cE_e}{eB\upsilon_{\rm out}^2} \nonumber \\
 &\sim 0.3\ {\rm yr} \left(\frac{\eta}{10}\right) \left(\frac{E_e}{1\ {\rm TeV}}\right)  \left(\frac{B}{79\ {\rm \mu G}}\right)^{-1} \left(\frac{\upsilon_{\rm out}}{0.1c}\right)^{-2} \\
t_{\rm dyn} &\sim  \frac{R_{\rm TS}}{\upsilon_{\rm out}}  \sim 0.7\ {\rm yr} \left(\frac{R_{\rm TS}}{7.1\times 10^{16}\ {\rm cm}}\right)  \left(\frac{\upsilon_{\rm out}}{0.1c}\right)^{-1}\\
t_{\rm syn} &\sim  \frac{6\pi m_e^2 c^3}{\sigma_{\rm T} B^2 E_e}  \sim 2\times 10^3\ {\rm yr}  \left(\frac{B}{79\ {\rm \mu G}}\right)^{-2} \left(\frac{E_e}{1\ {\rm TeV}}\right)^{-1},
\end{align}
where $e, m_e, E_e$ are respectively the electron's charge, mass, and energy, $\eta$ is the Bohm diffusion factor, and $\sigma_T$ is the Thomson cross section. Hence the maximum energy is almost always limited by the dynamical timescale rather than the synchrotron timescale. In this case,
\begin{align}
E_{\rm max} \sim 3\ {\rm TeV} \left(\frac{\eta}{10}\right)^{-1} \left(\frac{B}{79\ {\rm \mu G}}\right) \left(\frac{\upsilon_{\rm out}}{0.1c}\right) \left(\frac{R_{\rm TS}}{7.1\times 10^{16}\ {\rm cm}}\right),
\label{eq:E_max}
\end{align}
or
\begin{align}
\gamma_{\rm max} \sim 5\times 10^6 \left(\frac{\eta}{10}\right)^{-1} \left(\frac{B}{79\ {\mu G}}\right) \left(\frac{\upsilon_{\rm out}}{0.1c}\right) \left(\frac{R_{\rm TS}}{7.1\times 10^{16}\ {\rm cm}}\right).
\end{align}
Thus the average electron energy density in equation \eqref{eq:eps_e} is
\begin{align}
\langle e_e \rangle  \sim \int_{\gamma_{\rm min}}^{\gamma_{\rm max}} (\gamma_e m_ec^2) n(\gamma_e)\mathrm{d}\gamma_e  \sim fn_0 m_ec^2,
\label{eq:e_e}
\end{align}
where
\begin{align}
f = \int_{\gamma_{\rm min}}^{\gamma_{\rm max}} \gamma_e^{1-q} \mathrm{d}\gamma_e.
\end{align}
If we vary $q$ from $2$ to $3$, the factor $f$ changes from $\sim\log(\gamma_{\rm max}/\gamma_{\rm min})$ to $\sim\gamma_{\rm min}^{-1}$, which are from $\sim 14$ to $\sim 0.2$ in the case of $\gamma_{\rm min}\sim 5$ and $\gamma_{\rm max}\sim 5\times 10^6$. From equation \eqref{eq:eps_e} we obtain
\begin{align}
n_0 & \sim \frac{\epsilon_e L_{\rm out}}{(V_{\rm sh}/R_{\rm TS})\upsilon_{\rm out} f m_ec^2} \nonumber \\
&\sim 3.1\times 10^{-4}\ {\rm cm^{-3}} \nonumber \\
&\times  f^{-1}\left(\frac{\epsilon_e}{0.1}\right) \left(\frac{\rho}{10^3\ \mathrm{cm^{-3}}\ m_p}\right) \left(\frac{\upsilon}{10\ {\rm km\ s^{-1}}}\right)^{2}  \left(\frac{\upsilon_{\rm out}}{0.1c}\right).
\end{align}
Thus the number of electrons that become accelerated is a very small fraction of the entire electrons available ($\sim\rho/m_p$). The bulk of the electrons constitute a thermal population, probably obeying the Maxwell-Boltzmann statistics up to a trans-relativistic velocity. In the following calculations we neglect the contribution from thermal electrons, because with the magnetic field considered (equation \ref{eq:magnetic_field}) only non-thermal relativistic electrons can radiate in the GHz band via synchrotron (see next subsection).

\subsection{Radio Synchrotron Emission}
The typical frequency of the synchrotron emission is given as a function of the electron Lorentz factor $\gamma_e$ by
\begin{align}
\nu_{\rm syn}(\gamma_e) \sim \frac{eB}{2\pi m_e c} \gamma_e^2 \sim 1.4\ {\rm GHz} \left(\frac{B}{79\ {\rm \mu G}}\right) \left(\frac{\gamma_e}{2500}\right)^2,
\end{align}
which means that the synchrotron emission from non-thermal electrons is observed in the GHz band if they could be accelerated up to $\sim $ GeV energies. From equation \eqref{eq:E_max}, this condition is easily satisfied in the present case. The formula for the synchrotron spectrum from electrons with a power-law energy distribution is given by \citep{RadiPro}
\begin{align}
P_{\nu, \rm syn}  =  \frac{\sqrt{3}e^3n_0B\sin\alpha}{2\pi m_ec^2(q+1)}\Gamma\left(\frac{3q+19}{12}\right)\Gamma\left(\frac{3q-1}{12}\right)\left(\frac{2\pi m_ec\nu}{3eB\sin\alpha}\right)^{-(q-1)/2}.
\label{eq:synchrotron_spectrum}
\end{align}
Here $\alpha$ is the pitch angle and $\Gamma(x)$ is the Gamma function. For example, assuming a relatively hard spectrum of $q=2.2$, the synchrotron luminosity per unit volume, time, and frequency becomes
\begin{align}
P_{\nu, \rm syn}  
			 & \sim 3.3\times 10^{-35}\ {\rm erg\ s^{-1}\ cm^{-3}\ Hz^{-1}} \nonumber \\
			&\times  (\sin\alpha)^{1.6} \left(\frac{\epsilon_e}{0.1}\right)  \left(\frac{\epsilon_B}{0.1}\right)^{0.8}\left(\frac{\upsilon}{10\ {\rm km\ s^{-1}}}\right)^{3.6} \nonumber \\
			& \times \left(\frac{\rho}{10^3\ \mathrm{cm^{-3}}\ m_p}\right)^{1.8} \left(\frac{\upsilon_{\rm out}}{0.1c}\right)  \left(\frac{\nu}{1\ {\rm GHz}}\right)^{-0.6}.
\label{eq:synchrotron_spectrum_p22}
\end{align}
Therefore, the flux can be estimated as
\begin{align}
F_{\nu} &\sim \frac{P_{\nu, \rm syn} V_{\rm sh}}{4\pi D^2} \nonumber \\
&\sim 0.65\ {\rm \mu Jy} \nonumber \\
&\times (\sin\alpha)^{1.6} \left(\frac{\epsilon_e}{0.1}\right)  \left(\frac{\epsilon_B}{0.1}\right)^{0.8}\left(\frac{\upsilon}{10\ {\rm km\ s^{-1}}}\right)^{-3.9}   \left(\frac{\rho}{10^3\ \mathrm{cm^{-3}}\ m_p}\right)^{1.8}\nonumber \\
&\times \left(\frac{\tilde{\lambda}}{0.03} \right)^{5/4}  \left(\frac{M_{\rm BH}}{10\ M_\odot}\right)^3 \left(\frac{r_{\rm in}/r_{\rm S}}{3}\right)^{-5/4} \left(\frac{D}{8\ {\rm kpc}}\right)^{-2}.
\label{eq:Fnu}
\end{align}
Here $D$ is the distance from the source to Earth, and $D\approx 8$ kpc corresponds to IBHs near the bulge region. A notable point is that the flux is very sensitive to $\upsilon$ and $\rho$, with dependence 
\begin{align}
F_{\nu}\propto \upsilon^{(q-10)/2} \rho^{(q+5)/4}.
\end{align}
Hence the IBHs detectable by radio surveys are likely to be in a relatively slow velocity with respect to the ISM ($\upsilon \gtrsim c_{\rm ISM}$) and/or in dense parts of the ISM. This is the same trend with IBHs detectable with X-ray observations \citep{Tsuna18}, which implies that multi-wavelength observations are meaningful for these sources.

The absorption coefficient of electrons with a power-law energy distribution can be calculated by the following formula \citep{RadiPro}
\begin{align}
\alpha(\nu) &= \frac{\sqrt{3}e^3}{8\pi m_e} \left(\frac{3e}{2\pi m_e^3 c^5}\right)^{q/2} \nonumber \\
&n_0 (m_ec^2)^q (B\sin\alpha)^{(q+2)/2} \Gamma\left(\frac{3q+2}{12}\right) \Gamma\left(\frac{3q+22}{12}\right) \nu^{-(q+4)/2}.
\end{align}
Substituting $q=2.2$ for example, the optical depth due to self-absorption is
\begin{align}
\tau &\sim \alpha R_{\rm TS} \nonumber \\
& \sim 1\times 10^{-18}\left(\frac{R_{\rm TS}}{7.1\times 10^{16}\ {\rm cm}}\right)(\sin\alpha)^{2.1} \nonumber \\
&\times  \left(\frac{n_0}{3\times 10^{-4}\ {\rm cm^{-3}}}\right) \left(\frac{B}{79\ {\rm \mu G}}\right)^{2.1} \left(\frac{\nu}{1\ {\rm GHz}}\right)^{-3.1} \ll 1.
\end{align}
Thus we can neglect the effect of self-absorption as long as we consider the emission in the GHz band.

\subsection{Distribution of IBHs and ISM}
We use the same spatial, velocity, and mass distributions of IBHs previously modelled in \cite{Tsuna18}. We assume the total number of IBHs born in the past in the Galaxy to be $10^8$, roughly consistent with past estimates \citep[e.g.][]{Shapiro83,vandenHeuvel92,Samland98,Fender13,Caputo17}. The distribution of black holes at birth is assumed to have disc and bulge components, whose birth locations and histories are modelled to follow the observed stellar profiles in the Galaxy. Using cylindrical coordinates $(r,\theta,z)$, the black holes from the disc are modelled to form uniformly from 10 Gyrs ago to now, with a radial profile of $\propto \exp(-r/r_d)$ with $r_d = 2.15$ kpc \citep{Licquia15}, and uniformly in $|z|<h$, where $h=75$ pc is the scale height of molecular clouds in our Galaxy. The black holes from the bulge are modelled to form uniformly from 10 to 8 Gyrs ago, with a spherical exponential profile of $\propto \exp(-R/R_b)$ (where $R=\sqrt{r^2+z^2}$ and $R_b=120$ pc; \citealt{Sofue13}).

We calculate the initial velocity of each BH by the sum of the progenitor star's velocity and the natal kick. We assume progenitors from the disc follows the Galactic rotation curve approximated as
\begin{align}
\upsilon_\theta=
\begin{cases}
265-1875(r-0.2)^2&\mathrm{km\ s^{-1}}\ \ \ ($for$\  r<0.2)\\
225+15.625(r-1.8)^2&\mathrm{km\ s^{-1}}\ \ \ ($for$\  0.2<r<1.8)\\
225+3.75(r-1.8)&\mathrm{km\ s^{-1}}\ \ \ ($for$\  1.8<r<5.8)\\
240&\mathrm{km\ s^{-1}} \ \ \ ($for$\  r>5.8),
\end{cases}
\label{eq:rotation_curve}
\end{align}
where $r$ is in kpc. The progenitor stars from the bulge are assumed to have a Maxwell-Boltzmann distribution with mean $130\ {\rm km\ s^{-1}}$ \citep{Kunder12}. The natal kick velocity given upon a BH at its birth is assumed to have a Maxwell-Boltzmann distribution with its mean speed $\upsilon_{\rm avg}$ varying from $50\ {\rm km\ s^{-1}}$ to $400\ {\rm km\ s^{-1}}$. The spatial and velocity distributions of the IBHs are then obtained by a fourth-order Runge-Kutta calculation of the orbits of each BH under the Galactic potential model of \cite{Irrgang13} (their Model II). The mass distribution of black holes is assumed to be a Gaussian profile of mean $7.8\ \mathrm{M_{\sun}}$ and standard deviation $1.2\ \mathrm{M_{\sun}}$ \citep{Ozel10}.

We also use the same setting for the ISM as \cite{Tsuna18} that assumed to have five phases \citep{BH00}, which have different scale heights, densities and sound speeds (Table \ref{tab:ISMtable}; \citealp[see also][]{Agol02,Ioka17,Tsuna18}). The gas particle densities of the densest two phases, molecular clouds and cold H\,{\sevensize I}, are assumed to obey a power-law of index 2.8 and 3.8 respectively, with a range of $10^2\ {\rm cm^{-3}}\leq n \leq 10^5\ {\rm cm^{-3}}$ and $10\ {\rm cm^{-3}}\leq n \leq 10^2\ {\rm cm^{-3}}$ respectively. The distribution of the filling fraction of each phase in the Galaxy is calibrated with the observed surface density profile of $\mathrm{H_2}$ and H\,{\sevensize I} gases from \citet{NS16}. We assume an average particle mass of $2.72m_p$ for molecular clouds and $1.36m_p$ for atomic clouds. The effective sound speed of each phase, that includes the turbulent velocity which is important in cold phases, is set to $c_s=3.7(n/100\ {\rm cm}^{-3})^{-0.35}$ km\ s$^{-1}$ for molecular clouds \citep{MT05}, $c_s=150$ km\ s$^{-1}$ for the hottest H\,{\sevensize II} phase, and $10$ km\ s$^{-1}$ for the other three phases \citep{Ioka17}.

\begin{table*}
\centering
\begin{tabular}{ccccccc}
\hline 
Phase & $n_1\mathrm{[cm^{-3}]}$ & $n_2\mathrm{[cm^{-3}]}$ & $\beta$ & $H_d$ & $c_s[\mathrm{km\ s^{-1}}]$  & $\xi(r=8.3{\rm \ kpc})$ 
 \\ \hline \hline
Molecular clouds & $10^2$ & $10^5$ & $2.8$ & $75$ pc & $3.7(n/100\ {\rm cm}^{-3})^{-0.35}$ & $0.0004$ \\
Cold H\,{\sevensize I} & $10^1$ & $10^2$ & $3.8$ &  $150$ pc & $10$ & $0.026$ \\
Warm H\,{\sevensize I} & \multicolumn{2}{|c|}{$0.3$} & -- &  $500$ pc & $10$ & $0.46$ \\
Warm H\,{\sevensize II} & \multicolumn{2}{|c|}{$0.15$} & -- & $1$ kpc & $10$ & $0.16$  \\
Hot H\,{\sevensize II} & \multicolumn{2}{|c|}{$0.002$} & -- & $3$ kpc & $150$ & $0.37 $ \\ \hline
\end{tabular}
\caption{Summary of the ISM phases adopted in this work. We assume a power-law distribution for the number density for molecular clouds and cold H\,{\sevensize I} in the range $n_1 < n < n_2$ with an index $\beta$, but assume a single density for the other three phases. The parameters $H_d$ and $c_s$ are the disc scale heights \citep{Agol02} and effective sound speeds \citep{MT05,Ioka17} respectively. The filling fraction $\xi$ of each phase obtained from our modelling depends on the location in the Galaxy (see \citealp{Tsuna18} for details), and here the values of $\xi$ at the Sun's location ($r=8.3$ kpc) are shown.}
 \label{tab:ISMtable}
\end{table*}


\section{Results}
\label{sec:results}
We present the results on the number of IBHs detectable by future radio observations. As it is clear from the previous section, the radio flux and detectability depends on a number of model parameters. We first show the results using the parameters defined in Table \ref{table:parameters}, which we define as the ``optimistic" case. Then the dependence of our results on the model parameters is studied.

We consider the upcoming SKA1-mid, whose expected sensitivity is $A_{\rm eff}/T_{\rm sys} \approx 2\times 10^{3} \ {\rm m^2\ K^{-1}}$ at $950$ -- $1760$ MHz, and the planned SKA2 whose sensitivity is expected to be about an order of magnitude higher than SKA1 at the same frequency band. Here $A_{\rm eff}$ and $T_{\rm sys}$ are the effective area and the system temperature respectively. We adopt the expected sensitivity in the SKA All Sky Mid-Frequency Continuum Survey \citep{Norris15,Prandoni15}, and set the $5\sigma$ flux limit to $20\ {\rm \mu Jy}$ for SKA1-mid and $0.5\ {\rm \mu Jy}$ for SKA2 respectively. We compare this sensitivity with the source flux at 1 GHz obtained for each IBH (equation \eqref{eq:synchrotron_spectrum}, averaged over pitch angle $\alpha$), and regard the source as detectable if the source flux is larger than the sensitivity. We assume the Earth to be on the Galactic plane ($z=0$) with $r=8.3$ kpc, which is consistent with the Galactic potential we adopt from \cite{Irrgang13}.

\begin{table}
\centering
\begin{tabular}{l | c r}
  \hline
  \underline{Parameters on Accretion Flow} & & \\
  Parameter characterizing outflow in equation \eqref{eq:lambda_tilde} ($\tilde{\lambda}$)& & $0.05$\\
  Innermost radius of accretion disc ($r_{\rm in}$)& & 3$r_{\rm S}$ \\
  \underline{Parameters on Shock Microphysics} & & \\
  Magnetic-field amplification efficiency ($\epsilon_B$)& & 0.1\\
  Energy fraction given to non-thermal electrons ($\epsilon_e$)& & 0.1\\
  Power-law index of non-thermal electrons ($q$) & & 2 \\
  Bohm diffusion factor ($\eta$) & & 10 \\
  \hline
  \end{tabular}
   \caption{Model parameters used for our ``optimistic'' case.}
   \label{table:parameters}
\end{table}

\begin{figure}
 \begin{center}
  \includegraphics[width=1.0\linewidth]{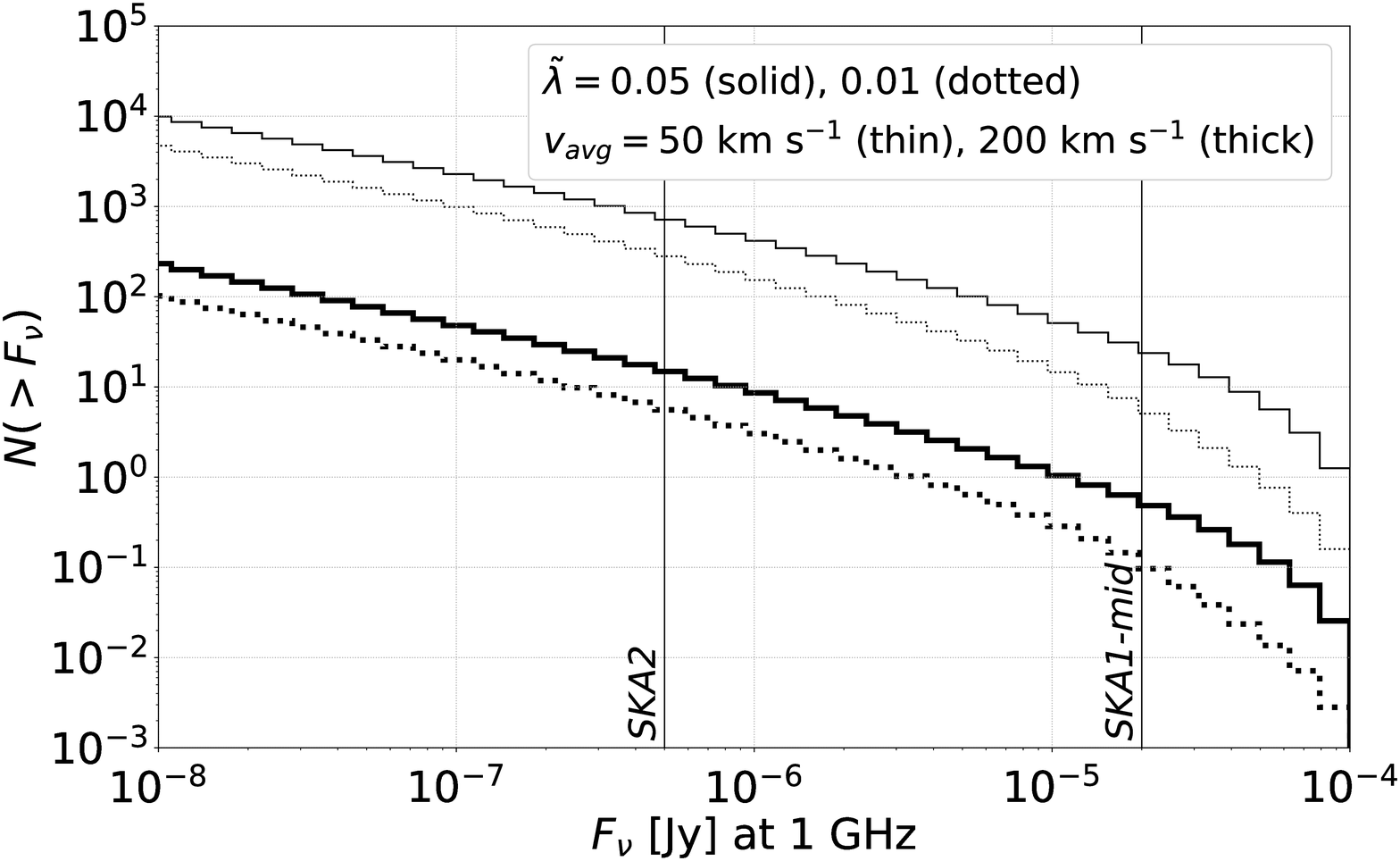} 
 \end{center}
\caption{Cumulative source counts of IBHs with radio flux (at 1 GHz) greater than $F_{\nu}$. The vertical lines show the expected sensitivity of SKA 1-mid and SKA2 in the mid-frequency band ($\sim 1$ GHz). We have assumed the parameters listed in Table \ref{table:parameters}, except for $\tilde{\lambda}$ which we show two values $0.01$ and $0.05$.}
\label{fig:flux_vs_number}
\end{figure}

Figure \ref{fig:flux_vs_number} shows the cumulative radio source counts, for a few different values of the average kick magnitude and parameter $\tilde{\lambda}$. We see that for our most optimistic parameter sets of $\tilde{\lambda}=0.05$ and $\upsilon_{\rm avg}=50\ {\rm km \ s^{-1}}$, about $30$ IBHs are expected to have radio flux observable by SKA1-mid. The number increases to $\sim 7\times 10^2$ for the case of SKA2.

\begin{figure}
 \begin{center}
  \includegraphics[width=1.0\linewidth]{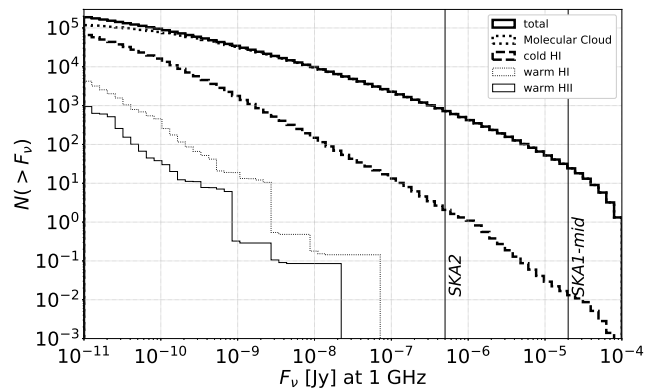} 
 \end{center}
\caption{Same as the $\tilde{\lambda}=0.05, \upsilon_{\rm avg}=50\ {\rm km\ s^{-1}}$ in Figure \ref{fig:flux_vs_number}, but broken into contributions from each ISM phase. The flux of IBHs residing in the hot HII phase is too low to appear in this figure.}
\label{fig:IBH_phases}
\end{figure}

Figure \ref{fig:IBH_phases} shows the contribution to the source counts from each ISM phase. As can be seen, the detectable IBHs are mostly hosted by molecular clouds instead of other ISM phases of smaller particle number density. We note that there may be intense star formation in a molecular cloud, and that ionizing photons emitted from young stars can create an H\,{\sevensize II} region, which can be bright in the GHz band via free-free emission, as is observed from the giant molecular cloud near the Galactic Centre, Sgr B2 \citep[e.g.,][]{Protheroe08}. Thus we expect that we can detect only IBHs whose host molecular clouds are not in the process of star formation. The intensity of star formation in nearby molecular clouds is observationally studied in \cite{Lada10}, and they derive that it is negligible for molecular clouds whose particle number density is lower than $\sim 10^{4}\ {\rm cm^{-3}}$. We find from our calculations that the fraction of radio-bright IBHs that reside in molecular clouds with number density higher than $\sim 10^4~{\rm cm}^{-3}$ (thus difficult to detect) is $\sim 10$ per cent for SKA1 and $\sim 3$ per cent for SKA 2, for our most optimistic parameter set of $\tilde{\lambda}=0.05$ and $\upsilon_{\rm avg}=50\ {\rm km \ s^{-1}}$. Although these numbers seem almost negligible, the fraction can be higher for more pessimistic parameter sets that would require higher particle number densities for detectable emission.

\begin{figure}
 \begin{center}
  \includegraphics[width=1.0\linewidth]{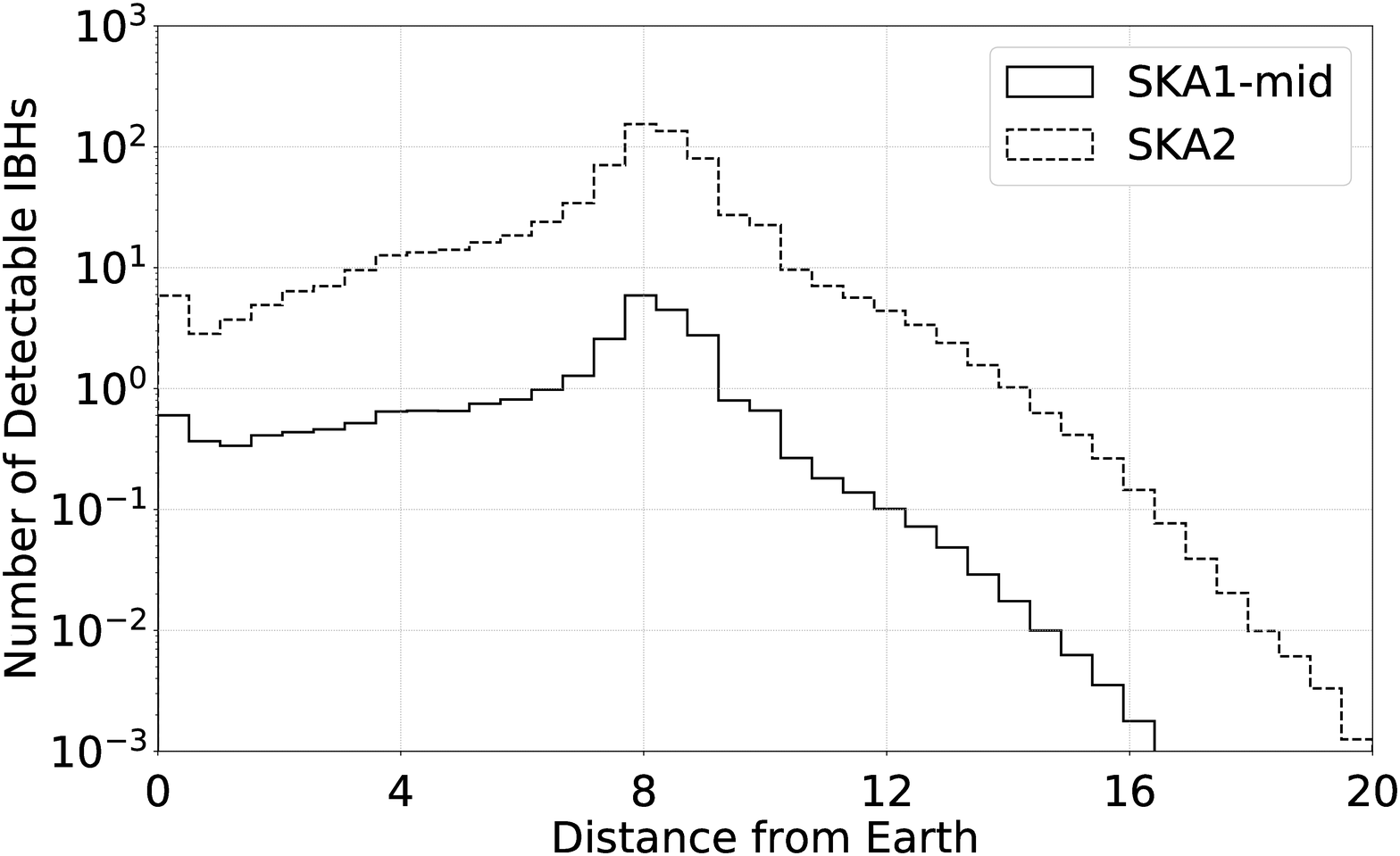} 
 \end{center}
\caption{Distribution of distance from Earth to IBHs detectable by all sky surveys with the two phases of SKA. For both of the distributions, we have assumed $\upsilon_{\rm avg}=50\ {\rm km\ s^{-1}}$ and the parameters listed in Table \ref{table:parameters}.}
\label{fig:distance_vs_number}
\end{figure}

We also study the distance of the IBHs detectable by the two SKA phases. We show the results for the case of our most optimistic parameter sets in Figure \ref{fig:distance_vs_number}. We find that the detectable IBHs are located near the Galactic Centre, due to the concentration of IBHs and the molecular clouds around the Galactic Centre (see \citealt{Tsuna18} for details). This shows that potential SKA surveys targeting IBHs can be more optimized by allocating more exposure time around the Galactic Centre. For example, in the case of SKA1-mid, a deeper survey focusing on $\sim 100$ deg$^2$ around the Galactic Centre can enhance the sensitivity by an order of magnitude (\citealp{Prandoni15}; see their figure 1 bottom panel), and can increase the number of detections by up to an order of magnitude (see Fig. \ref{fig:flux_vs_number}).

\begin{figure*}
 \begin{tabular}{cc}
 
 \begin{minipage}{0.5\hsize}
 \centering
\includegraphics[width=0.95\linewidth]{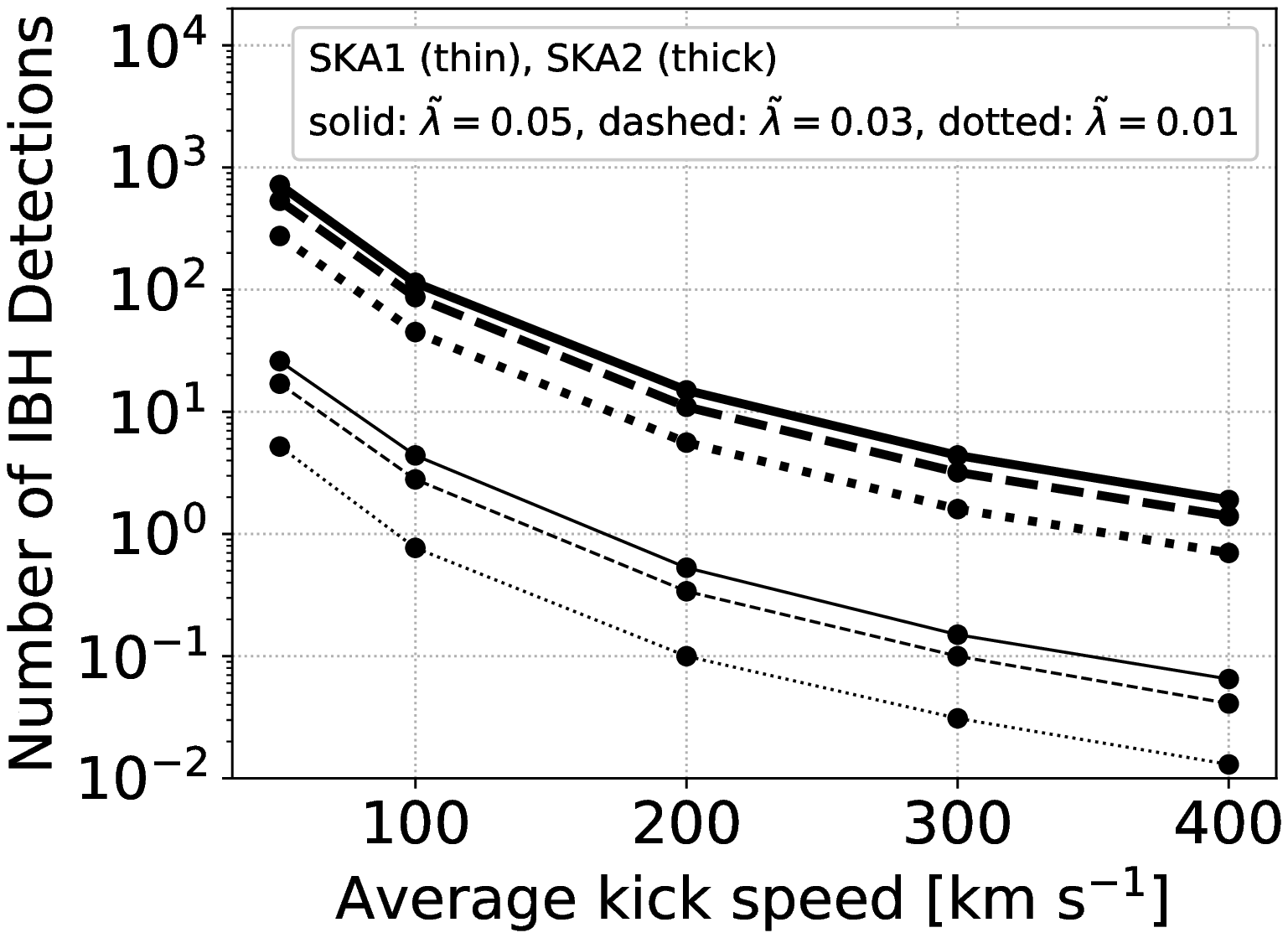} 
 \caption{Expected number of IBH detections by SKA1-mid and SKA2 for various mean values of kick magnitude, varying $\tilde{\lambda}$ as a free parameter. We set other parameters the same as in Table \ref{table:parameters}.}
\label{fig:lambda_tilde}
\end{minipage}

\begin{minipage}{0.5\hsize}
 \centering
\includegraphics[width=0.95\linewidth]{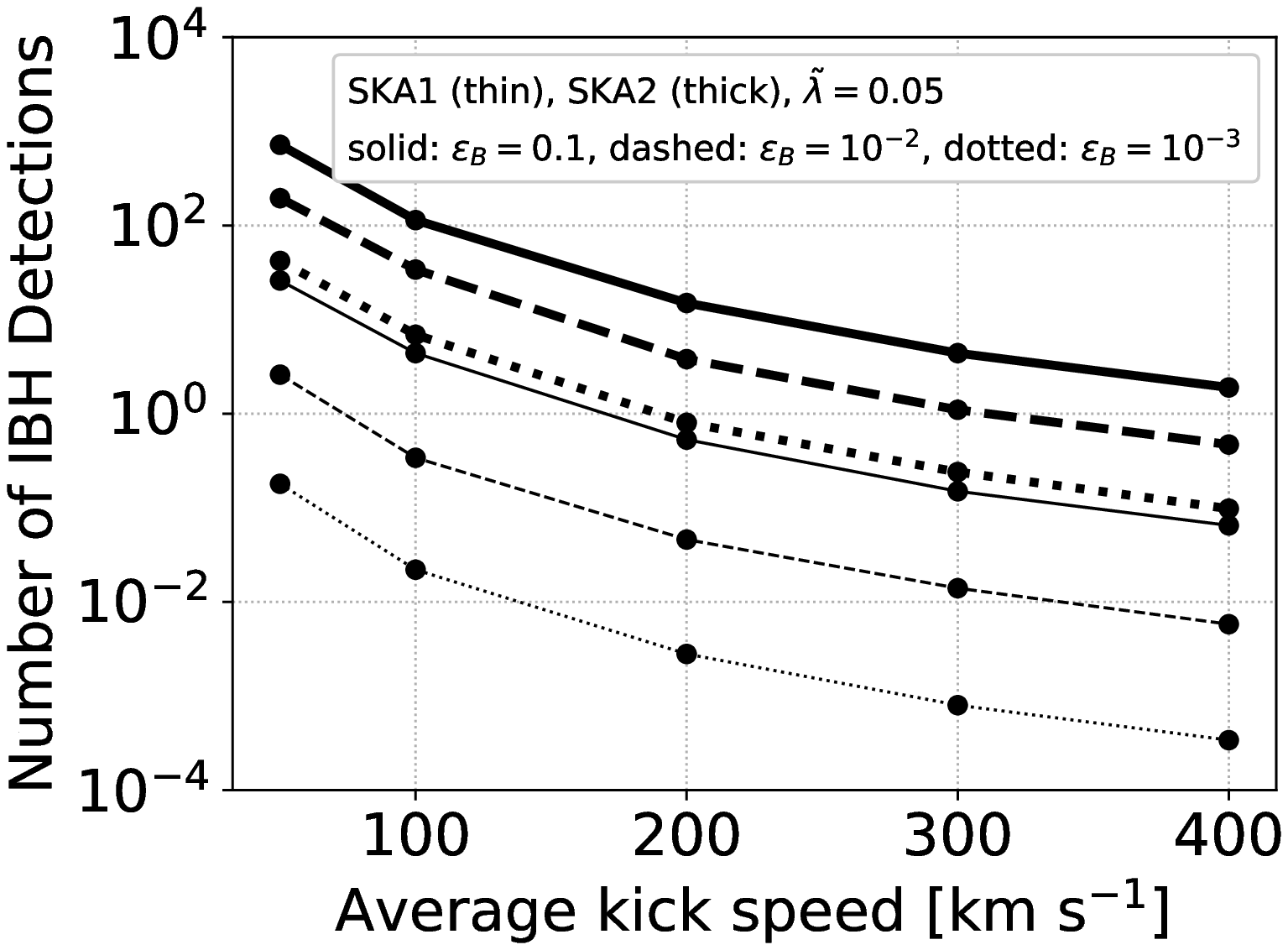}
 \caption{Same as Figure \ref{fig:lambda_tilde}, but for the case of varying only $\epsilon_B$.}
\label{fig:epsB}
\end{minipage}

\end{tabular}
\end{figure*}

\begin{figure*}
 \begin{tabular}{cc}
 
 \begin{minipage}{0.5\hsize}
\centering
\includegraphics[width=1.0\linewidth]{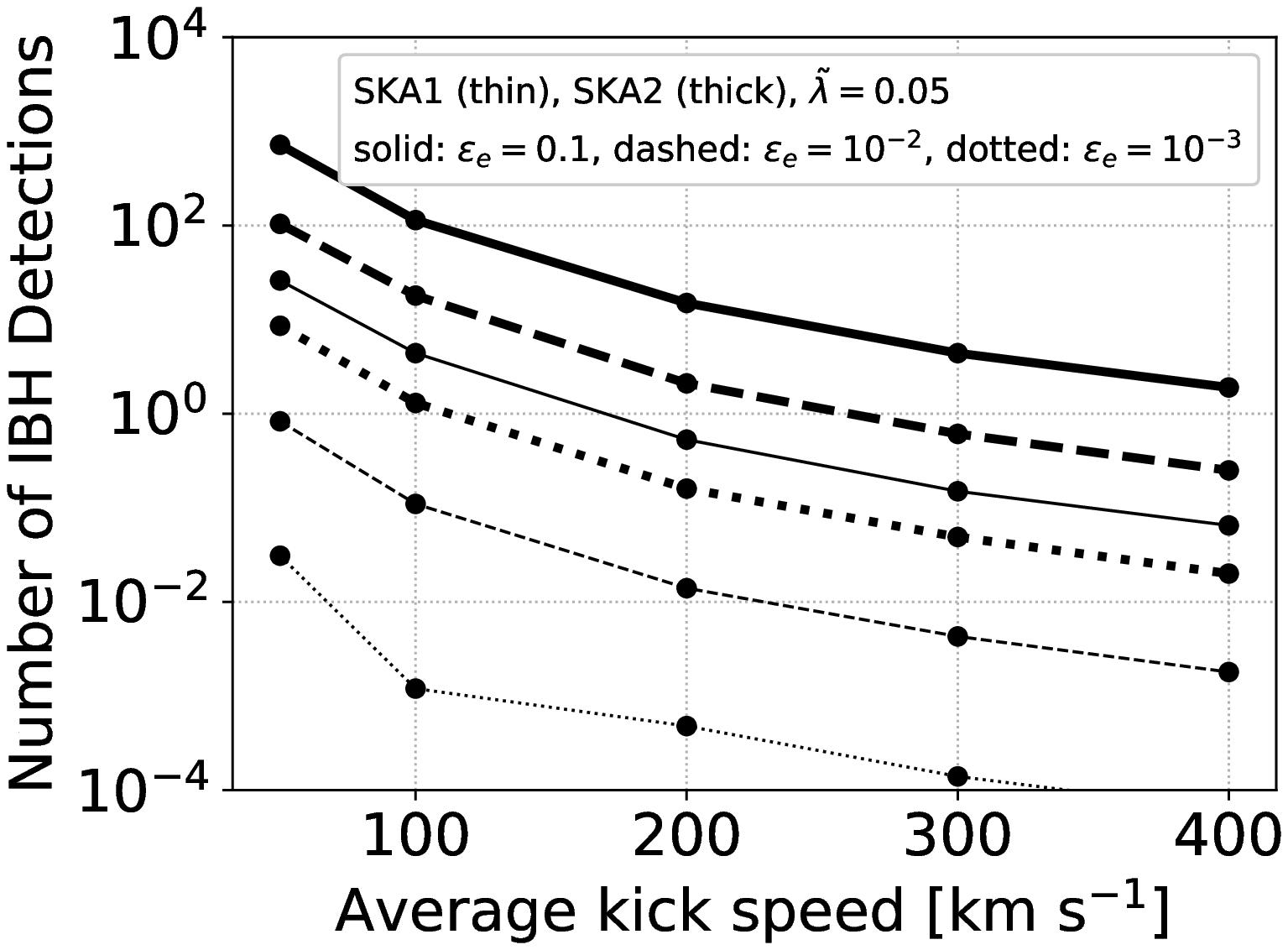}
 \caption{Same as Figure \ref{fig:lambda_tilde}, but for the case of varying only $\epsilon_e$.}
\label{fig:epse}
 \end{minipage}
 
\begin{minipage}{0.5\hsize}
 \centering
\includegraphics[width=1.0\linewidth]{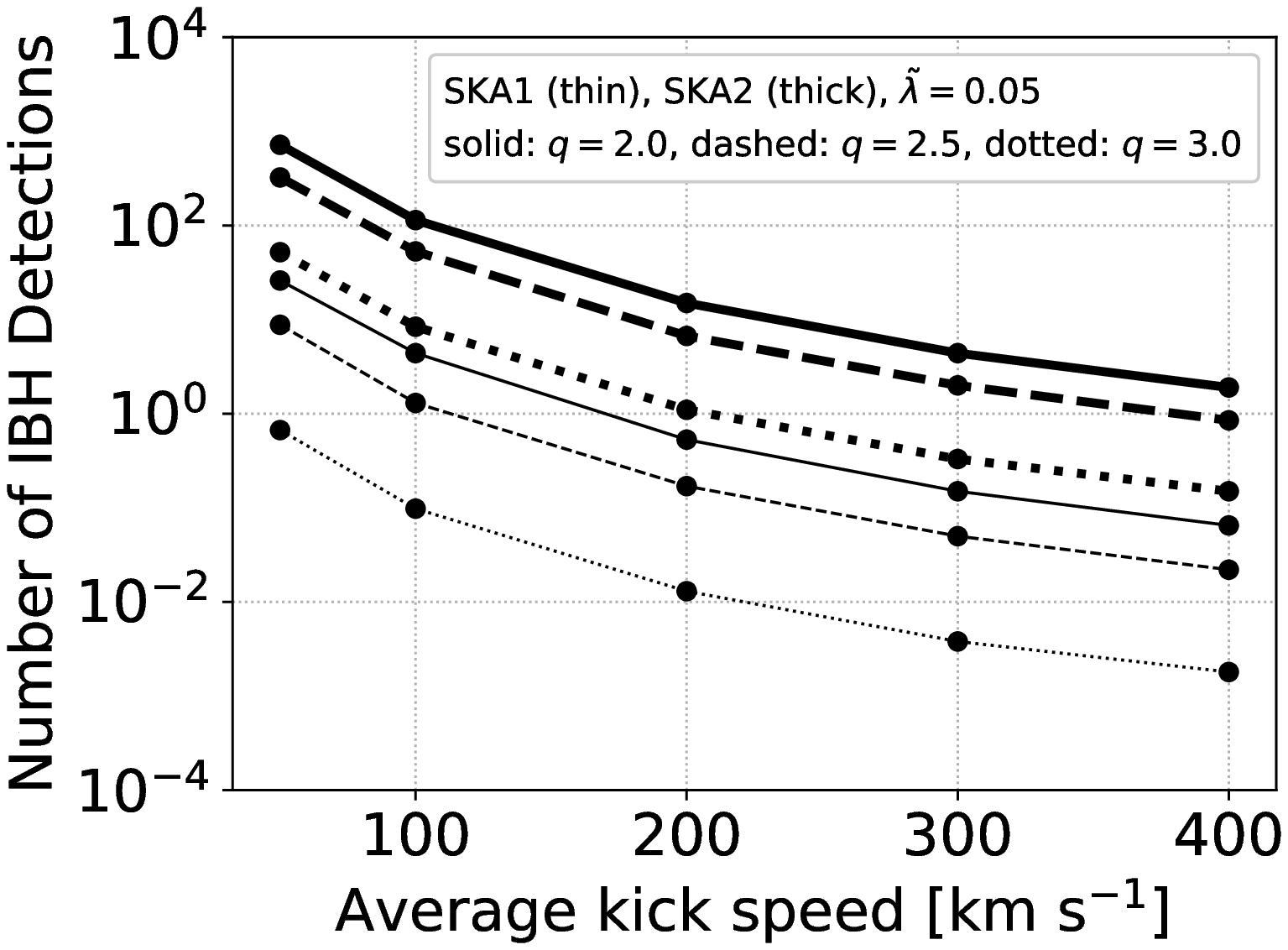}
 \caption{Same as Figure \ref{fig:lambda_tilde}, but for the case of varying only $q$.}
\label{fig:q}
\end{minipage}

\end{tabular}
\end{figure*}

Among the parameters listed in Table \ref{table:parameters}, the uncertain parameters that can greatly alter the results are $\tilde{\lambda}, \epsilon_B, \epsilon_e$, and $q$. To see the dependence of our results on these four parameters, we calculate the number of detections by varying each parameter, while keeping the other parameters same as in Table \ref{table:parameters}. The results are shown in Figures \ref{fig:lambda_tilde} -- \ref{fig:q}. The number of detectable sources is reduced by almost three orders of magnitude when considering the most pessimistic value of $\epsilon_e$, by roughly two orders of magnitude when for the most pessimistic $\epsilon_B$ or $q$, and by at most an order of magnitude for the most pessimistic $\tilde{\lambda}$. Assuming values of the phenomenological parameters $\epsilon_e$ and $\epsilon_B$ implied from other observations, e.g. $\epsilon_e\sim0.1, \epsilon_B\sim 0.01$ obtained from observations of afterglows of gamma-ray bursts \citep[e.g.][]{Meszaros06} or $\epsilon_e\sim0.01, \epsilon_B\sim 0.1$ obtained from radio and X-ray observations of Type IIb supernovae \citep[e.g.][]{Maeda12}, will both reduce the number of detectable IBHs by about an order of magnitude. 

\section{Discussion}
\label{sec:discussion}

\subsection{Caveats}
In this paper we consider a power-law model of the accretion disc, which is a very simplified approximation. Recent numerical works that solve the structure of RIAFs predict that the accretion flow has multiple components, where convection dominates in the inner region \citep{Inayoshi18}. The resulting fraction of accreted matter that actually reaches a BH will be as small as $10^{-3}\lesssim \lambda \lesssim 10^{-2}$, and moreover the contribution from the fast outflow close to the BH becomes much smaller. These factors may lead to inefficient electron acceleration and radio emission. 

This work put an upper bound on the mass accretion rate as $\dot{M}<0.1\dot{M}_{\rm Edd}$ for the BHs to be radio sources, assuming that only RIAFs are possible to generate this outflow and radio emission. The standard discs \citep{Shakura73} which have higher accretion rates than RIAFs are considered to not generate this kind of outflow due to efficient radiative cooling, but a slim disc model \citep{Abramowicz88}, which have even higher accretion rates reaching the Eddington limit, has been considered to generate outflows \citep{Shakura73,Meier82}. However, the fraction of IBHs that reach this huge amount of accretion rate is negligible.

\subsection{Required Dynamic Range}
In this work we have found that the detectable sources mainly reside near the Galactic Centre, with a typical distance of $0.1$--$1$ kpc. This distance corresponds to an angular separation of roughly $0.7$--$7$ degrees, for a source distance of $\sim 8$ kpc. Thus with SKA's field of view of 1 deg$^2$, there is a non-negligible possibility that the sources are contained in the same field of view as Sgr A*, a bright radio source in the Galactic Centre. This possible high contrast gives a requirement on the dynamic range to detect the faint IBH.

The radio flux from Sgr A* is $\sim 0.6$ Jy at $1$ GHz \citep{Falcke98}, and the typical flux at 1 GHz of a strongest radio source outside Sgr A* is $\sim 0.16$ Jy\footnote{\url{https://www.skatelescope.org/uploaded/12336_114_Memo_Condon.pdf}}. Using the aforementioned sensitivities of SKA1-mid ($20\ \mu$Jy) and SKA2 ($0.5\ \mu$Jy), we find that the required dynamic ranges to detect the faintest IBHs detectable by the SKA1-mid and SKA2 are $45$ dB and $61$ dB respectively for IBHs overlapping with Sgr A*, and $39$ dB and $55$ dB respectively for IBHs not overlapping with Sgr A*.

\subsection{Foreground Emission from the Galactic Centre} 
The Galactic Centre is prominent in diffuse synchrotron foreground emission, which may be a problem for observing faint radio sources. However, this diffuse emission is in principle distinguishable from point sources like what we consider here, as interferometers like SKA measures the Fourier components of the brightness map. The fluctuation of the synchrotron emission would be observable by SKA and may become a contamination. Observing at higher frequencies than 1 GHz may help to mitigate this, as the foreground would be weaker and the angular resolution of SKA would be better. The flux from the IBH will be weaker at higher frequencies as well, with frequency dependence $\nu^{-(q-1)/2}$. Thus, it may be a better strategy to observe at higher frequencies if the electrons can be efficiently accelerated ($q\sim 2$). The detailed evaluation on the observational strategies and their feasibility is beyond the scope of this paper, and we leave quantitative discussions to future work.

\subsection{Possible Constraints on BH Natal Kicks}
Little is known about the natal kicks of BHs and its mechanism \citep[see][for a review]{Belczynski16}. Some works assume that natal kicks are inversely proportional to the mass of the BHs. This implies a natal kick velocity much lower than that of neutron stars (of typically $\sim 400 \ {\rm km \ s^{-1}}$; \citealp{Hobbs05}). However, other works \citep{Repetto12,Repetto15,Repetto17} claim that a fraction of the BHs require natal kicks of similar magnitude as neutron stars to reproduce the observed distribution of X-ray binaries in our Galaxy.

A non-detection of radio-emitting IBH candidates allows us to constrain the natal kicks, as the number of expected detections varies by almost 3 orders of magnitude within the range of $\upsilon_{\rm avg}$ we considered. For example, if we assume $\epsilon_e=0.01$, $\epsilon_B=0.1$, $q=2$ and $\tilde{\lambda}=0.05$, a future non-detection by SKA2 will constrain $\upsilon_{\rm avg}$ to be $\upsilon_{\rm avg}\gtrsim 200\ {\rm km\ s^{-1}}$ (see Figure \ref{fig:epse}). A statistical constraint on BH natal kicks like this may give important implications on its mechanism.

\subsection{Radio and Multi-Wavelength Follow-up}
There are other kinds of sources that SKA can detect, including extragalactic sources such as active galactic nuclei and starburst galaxies. The angular resolution of SKA with a space-based VLBI ($\sim 100\ {\rm \mu as}$ for a source of brightness $\sim 1\ \mu$Jy; \citealp{Taylor08}) may help discriminate IBHs from these extragalactic sources. Isolated BHs near the Galactic Centre (distance of $\sim 8$ kpc) moving at a relative velocity of $\sim 10\ {\rm km\ s^{-1}}$ can have an offset as large as $\sim 250\ {\rm \mu as\ yr^{-1}}$, and a parallax of $\sim 130\ {\rm \mu as\ yr^{-1}}$. These are detectable by follow-up observations, in contrast to extragalactic sources that will have much smaller proper motions and parallaxes. This idea of measuring their proper motions has been previously demonstrated to be feasible by \cite{Fender13} that considered the SKA detectability of more nearby BHs (within a distance $\lesssim 250$ pc). Our results, based on realistic modeling of the BHs and the interstellar gas, predict that the detectable IBHs are rather near the Galactic Centre, which is a different population from \cite{Fender13}. Although detecting the proper motions would be more difficult as it would be farther, the above simple estimation shows that it is possible to detect their proper motions by follow-up observations for at least a non-negligible fraction of detectable IBHs.

The high angular resolution may also be helpful to distinguish from X-ray binaries, in the case where the binary companion is too dim to be observed by optical or infrared telescopes. Furthermore, the detectability of the parallax can also help to obtain an accurate distance of the detectable IBHs. This will be a useful information, complementary to other probes, for knowing the Galactic distribution of black holes. 

The accurate distance of the source can also be helpful for multi-wavelength followup. The radio emission depicted here contains many uncertain model parameters, but there are expectations that IBHs accreting from molecular clouds may be bright enough to be observable by hard X-ray surveys as well \citep{Tsuna18}. The obtained distance can be used to estimate the luminosity, and thus the accretion rate of the IBH. If we are able to know the properties of the host molecular cloud (e.g. temperature, density) from radio observations \citep[as is done by CO observations in e.g.,][]{Sanders85}, it is possible to infer the mass of the BH. Although the inferred masses would contain significant uncertainties that possibly arise from e.g. the black hole velocity and the parameter $\lambda$, this can be an independent way to probe the mass function of stellar-mass black holes, which are so far only understood from binary systems.


\section{Conclusions}
\label{sec:conclusions}

If black holes accrete gas at accretion rates much smaller than the Eddington rate, outflows can be launched from the inner disc region. This condition is usually satisfied for accreting IBHs, that do not have companion stars but are considered to accrete material from the surrounding interstellar medium. In this work, we considered radio emission from the interaction of these outflows and the ISM, and studied its detectability by future radio observations. To obtain the radio luminosity distribution of IBHs, we have obtained the spatial and velocity distributions of IBHs through the calculation of the orbits.

We find that for our most optimistic parameters we can detect around 30 IBHs as radio sources by the upcoming SKA1-mid, and about 700 IBHs by the future SKA2. However we note that the results depend on the phenomenological parameters, $\epsilon_e$ and $\epsilon_B$, which are determined from the microphysical processes occurring in collisionless shocks. Lower values of these parameters can significantly reduce the number of detectable IBHs, possibly making an IBH detection difficult even by these future radio observations. 

One key advantage of radio observations is that we can accurately know the positions of the sources in the sky thanks to the high angular resolution. We find that distances of some of these detectable IBHs can be obtained via follow-up parallax measurements, which would be helpful in knowing the Galactic distribution of these black holes. Another significant point is that the characteristics of radio detectable IBHs and their host ISMs are similar to those of X-ray detectable IBHs studied in previous works \citep[e.g.,][]{Agol02,Fender13,Tsuna18}. This shows that, just like black holes in X-ray binaries, multi-wavelength observations are fruitful to further understand the properties of black holes in our Galaxy.

\section*{Acknowledgements}
The authors thank the anonymous referee for his/her comments that improved this manuscript to a great extent. The authors also thank Keitaro Takahashi, Kazumi Kashiyama, and Riouhei Nakatani for valuable discussions. DT is supported by the Advanced Leading Graduate Course for Photon Science (ALPS) at the University of Tokyo, and by JSPS KAKENHI Grant Number 19J21578. NK is supported by the Hakubi project at Kyoto University. 

\bibliographystyle{mnras} \bibliography{IBH}

\bsp	
\label{lastpage}
\end{document}